\documentclass[conference]{IEEEtran}
\usepackage{multirow}
\usepackage{graphicx}
\usepackage{amsmath}
\usepackage{color}
\usepackage{epsfig}
\usepackage{amsfonts}
\usepackage{amssymb}
\usepackage{amsthm}
\usepackage[usenames,dvipsnames]{pstricks}
\usepackage{epstopdf}
\usepackage{algorithm}
\usepackage{caption}
\usepackage{subcaption}
\usepackage{mathtools}
\usepackage[noend]{algpseudocode}

\usepackage{array}
\usepackage{multirow}
\usepackage{bm}
 
\begin{document}
\title{{\LARGE A DNN-based OTFS Transceiver with Delay-Doppler Channel Training and IQI Compensation}}
\author{Ashwitha Naikoti and A. Chockalingam  \\
\text{Department of ECE, Indian Institute of Science, Bangalore 560012}}

\maketitle

\begin{abstract}
In this paper, we present a deep neural network (DNN) based transceiver architecture for delay-Doppler (DD) channel training and detection of orthogonal time frequency space (OTFS) modulation signals along with IQ imbalance (IQI) compensation. The proposed transceiver learns the DD channel over a spatial coherence interval and detects the information symbols using a single DNN trained for this purpose at the receiver. The proposed transceiver also learns the IQ imbalances present in the transmitter and receiver and effectively compensates them. The transmit IQI compensation is realized using a single DNN at the transmitter which learns and provides a compensating modulation alphabet (to pre-rotate the modulation symbols before sending through the transmitter) without explicitly estimating the transmit gain and phase imbalances. The receive IQI imbalance compensation is realized using two DNNs at the receiver, one DNN for explicit estimation of receive gain and phase imbalances and another DNN for compensation. Simulation results show that the proposed DNN-based architecture provides very good performance, making it as a promising approach for the design of practical OTFS transceivers. 
\end{abstract}
{\em {\bfseries Keywords}} --
{\footnotesize {\em \small OTFS modulation, delay-Doppler domain, deep neural networks, channel training, signal detection, IQ imbalance.}}

\vspace{-1mm}
\section{Introduction}
\label{sec1}
\let\thefootnote\relax\footnotetext{This work was supported in part by the J. C. Bose National Fellowship, Department of Science and Technology, Government of India.} Next generation wireless communication systems are envisaged to provide reliable service at low latencies in high-mobility environments. Recently, orthogonal time frequency space (OTFS) modulation which achieves superior performance compared to OFDM in high-Doppler channels has been proposed \cite{b1}-\cite{b4}. OTFS is a two-dimensional modulation technique in which information symbols are multiplexed in the delay-Doppler (DD) domain \cite{b5}-\cite{b9}. The channel is also represented in the DD domain which converts the rapidly varying channel in the time domain into an almost time-invariant channel in the DD domain. In terms of implementation, OTFS can be implemented as an overlay on any multicarrier modulation scheme such as OFDM through the use of pre- and post processing blocks. Such attractive performance and implementation attributes of OTFS have created significantly growing interest in OTFS research. In particular, issues related to practical implementation of OTFS have gained importance. In this context, our focus in this paper is to address the following key issues in OTFS transceiver design: 1) channel   estimation and signal detection, and 2) IQ imbalance (IQI) due to hardware imperfections and its compensation at the transmitter and receiver. The approach we adopt in this effort is the use of deep neural networks (DNN).

The field of deep learning is rapidly growing and neural networks are increasingly being considered for various purposes in the design of intelligent communication systems. Deep learning based approaches have been used in solving communication problems such as constellation/code design, channel decoding, signal detection, and channel estimation \cite{b10}-\cite{b16}. Transceiver designs using direct conversion architecture in the RF front-end are becoming popular as they provide a highly integrated architecture at reduced cost \cite{b16a}. A key issue, however, is the need to deal with RF impairments \cite{b17},\cite{b17a}. It is a challenge to achieve perfect match between the in-phase (I) and quadrature-phase (Q) paths in a transceiver chain, more so at mmwave frequencies. Large gain and phase imbalances between I and Q paths can lead to severe performance degradation, particularly in direct conversion architectures, and techniques to compensate such impairments are important \cite{b19}-\cite{b21}. In addition to the need to address the issue of IQI, channel estimation overhead reduction has always been desired in communication systems design. 

In this paper, motivated by the above observations, we develop a unified DNN-based transceiver architecture that carries out DD channel training and detection of OTFS signals along with IQI compensation at the transmitter and receiver. The key highlights of the proposed DNN framework can be summarized as follows. 
\begin{itemize}
\item The proposed DNN-based transceiver learns the DD channel over a spatial coherence interval and detects the information symbols using a single DNN trained for this purpose at the receiver. 
\item The proposed transceiver also learns the IQ imbalances present in the transmitter and receiver and effectively compensates them. 
\item The transmit IQI compensation is realized using a single DNN at the transmitter which learns and provides a compensating modulation alphabet (to pre-rotate the modulation symbols before sending through the transmitter) without explicitly estimating the transmit gain and phase imbalances. 
\item The receive IQI imbalance compensation is realized using two DNNs at the receiver, one DNN for explicit estimation of receive gain and phase imbalances and another DNN for compensation. 
\end{itemize}
\begin{figure*}[t]
\centering
\includegraphics[width=15cm, height=1.65cm]{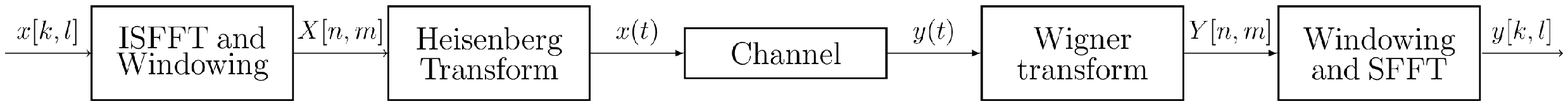}
\caption{OTFS modulation scheme.}
\label{fig1}
\end{figure*}
Our simulation results show that the proposed DNN-based architecture provides very good performance, making DNN approach as a promising approach for the design of practical OTFS transceivers.  
The rest of the paper is organized as follows. The considered OTFS system model in the presence of transmit and receive IQ imbalance is presented in Sec. \ref{sec2}. The proposed DNN architectures are presented in Sec. \ref{sec3}. Simulation results and discussions are presented in Sec. \ref{sec4}. Conclusions are presented in Sec. \ref{sec5}.

\section{OTFS System Model}
\label{sec2}
In this section, we introduce the OTFS system model and the vectorized input-output relation in the presence of IQ imbalance at transmitter and receiver. The block diagram of OTFS modulation is shown in Fig. \ref{fig1}.

\subsection{OTFS modulation}
\label{sec2a}
The OTFS transmitter multiplexes $NM$ information symbols from a modulation alphabet $\mathbb{A}$, denoted by $x[k,l]$, $k=0,\cdots,N-1$, $l=0,\cdots,M-1$ on a $N\times M$ DD grid. These symbols occupy a bandwidth of $M\Delta f$ and are transmitted in a time duration of $NT$, where $\Delta f=1/T$. The symbols in the DD domain are mapped to the time-frequency (TF) domain by inverse symplectic finite Fourier transform (ISFFT), as  
\begin{equation}
X[n,m]=\frac{1}{MN} \sum_{k=0}^{N-1} \sum_{l=0}^{M-1} x[k,l] e^{j2 \pi (\frac{nk}{N}-\frac{ml}{M})}. 
\label{eq1}
\end{equation}
The above TF domain signal is transformed into a time domain signal $x(t)$ using Heisenberg transform, as 
\begin{equation}
x(t)=\sum_{n=0}^{N-1} \sum_{m=0}^{M-1} X[n,m]g_{tx}(t-nT)e^{j2 \pi m \Delta f(t-nT)},
\label{eq2}
\end{equation}
where $g_{tx}(t)$ is the transmit pulse shape. The signal $x(t)$ is transmitted through a channel having a DD domain response $h(\tau,\nu)$, where $\tau$ and $\nu$ denote delay and Doppler variables. The received time domain signal $y(t)$ at the receiver is  
\begin{equation}
y(t)=\int_{\nu} \int_{\tau} h(\tau,\nu)x(t-\tau)e^{j2 \pi \nu(t-\tau)} d\tau d\nu.
\label{eq3} 
\end{equation}
The received signal $y(t)$ is transformed into a TF domain signal using Wigner transform, as 
\begin{equation}
Y[n,m]=A_{g_{rx},y}(t,f)|_{t=nT,f=m \Delta f},
\label{eq4}
\end{equation}
\begin{equation*}
A_{g_{rx},y}(t,f)=\int g^{*}_{rx}(t'-t)y(t)e^{-j2 \pi f(t'-t)}dt',
\end{equation*}
where $g_{rx}(t)$ is the receive pulse shape. If $g_{rx}(t)$ and $g_{tx}(t)$ satisfy the biorthogonality condition \cite{b2}, the following equation gives the input-output relation in the TF domain
\begin{equation}
Y[n,m]=H[n,m]X[n,m]+V[n,m],
\label{eq5}
\end{equation}
where $V[n,m]$ is noise in TF domain and $H[n,m]$ is 
\begin{equation}
H[n,m]=\int_{\tau} \int_{\nu} h(\tau,\nu)e^{j2 \pi \nu nT}e^{-j2 \pi (\nu + m \Delta f)\tau}d \nu d \tau.
\label{eq6}
\end{equation}
Applying symplectic finite Fourier transform (SFFT), the TF signal $Y[n,m]$ is mapped to the DD domain signal $y[k,l]$, as 
\begin{equation}
y[k,l]=\sum_{n=0}^{N-1} \sum_{m=0}^{M-1} Y[n,m]e^{-j2 \pi (\frac{nk}{N}-\frac{ml}{M})}.
\label{eq7}
\end{equation}
From (\ref{eq1})-(\ref{eq7}), the input-output relation can be written as \cite{b2}  
\begin{equation}
y[k,l]=\frac{1}{MN}\hspace{-0.5mm} \sum_{l'=0}^{N-1} \sum_{k'=0}^{M-1} \hspace{-0.5mm} x[k',l'] h_w(\frac{k-k'}{NT},\frac{l-l'}{M \Delta f}) + v[k,l],
\label{eq8}
\end{equation}
where $h_w(\nu,\tau)$ is the circular convolution of the channel response with a windowing function $w(\nu,\tau)$ and $h_w(\frac{k-k'}{NT},\frac{l-l'}{M \Delta f}) = h_w(\nu,\tau)|_{\nu =\frac{k-k'}{NT}, \tau = \frac{l-l'}{M \Delta f}}$. Now, consider a $P$-path DD channel with response 
\begin{equation}
h(\tau,\nu) = \sum_{i=1}^{P} h_i \delta(\tau-\tau_i) \delta(\nu-\nu_i),
\label{eq9}
\end{equation} 
where $h_i, \tau_i$, and $\nu_i$ are the channel gain, delay, and Doppler shift corresponding to the $i$th path, respectively. We assume $\tau_i \triangleq \frac{\alpha_i}{M \Delta f}$ and $\nu_i \triangleq \frac{\beta_i}{NT}$ where $\alpha_i, \beta_i$ are 
integers. For rectangular transmit and receive windowing functions,
the input-output relation in (\ref{eq8}) can be obtained as \cite{b6}:
\begin{equation}
y[k,l]=\sum_{i=1}^{P} h'_i x[(k-\beta_i)_N, (l-\alpha_i)_M] + v[k,l],
\label{eq10}
\end{equation}
where $h'_i = h_i e^{-j2 \pi \nu_i \tau_i}$ such that $h_i$s are i.i.d. $\mathcal{CN}(0,1/P)$. The above input-output relationship can be vectorized as \cite{b4}
\begin{equation}
\mathbf{y=Hx+v},
\label{eq11}
\end{equation}
where $\mathbf{x} \in \mathbb{C}^{MN \times 1}$ is the transmitted vector, $\mathbf{y} \in \mathbb{C}^{MN \times 1}$ is the received vector, $\mathbf{v} \in \mathbb{C}^{MN \times 1}$ is the noise vector, and $\mathbf{H} \in \mathbb{C}^{MN\times MN}$ is the DD domain effective channel matrix. The element $x[k,l]$ in the DD grid is the $(k + Nl)$th element in $\mathbf{x}, \; k=0,\cdots,N-1, l=0,\cdots,M-1$.

\begin{figure*}[t]
\centering
\includegraphics[width=17cm, height=8.5cm]{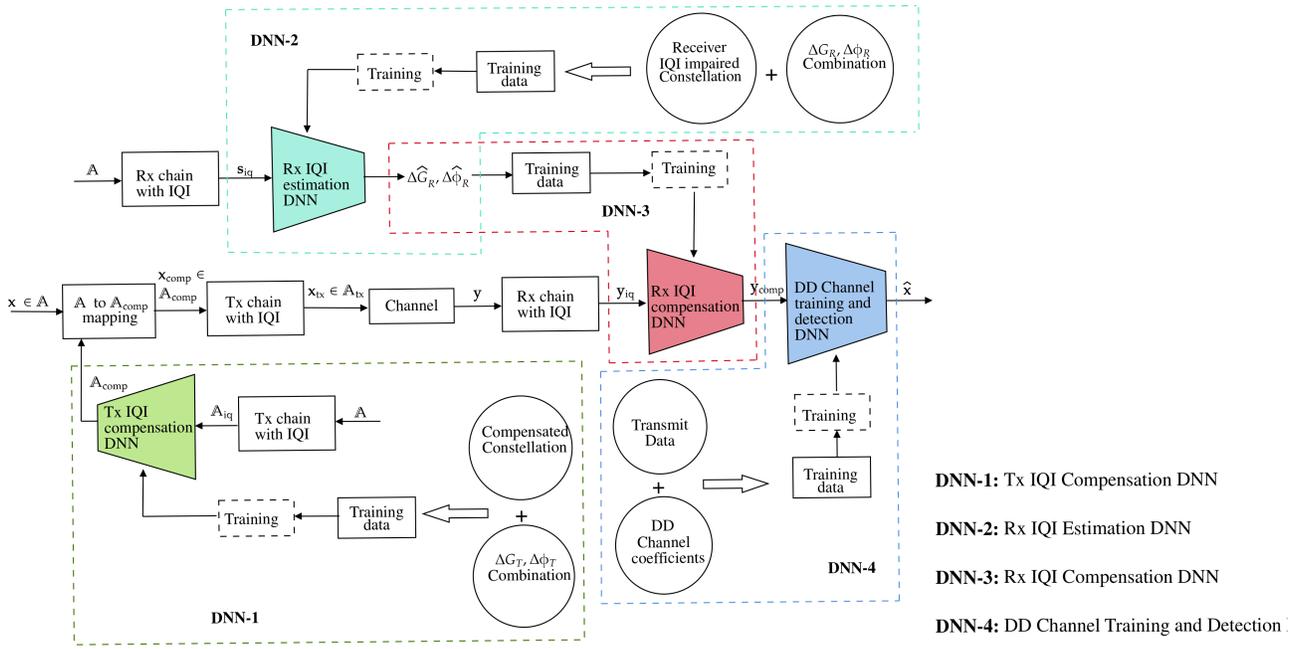}
\caption{Proposed DNN-based OTFS transceiver architecture.}
\label{fig2}
\vspace{-1mm}
\end{figure*}

\subsection{OTFS system model with IQI}
\label{sec2b}
The IQ imbalance in transceivers is modeled using two parameters, namely, gain imbalance ($\Delta G$) and phase imbalance ($\Delta\phi$) between the I-arm and Q-arm. 

{\em Transmitter IQI}: Consider a transmit RF chain which is impaired by IQI. Let $\Delta G_{T}$ and $\Delta \phi_{T}$ denote the gain and phase imbalances, respectively, at the transmitter. The transmit vector with IQI can be written as \cite{b17},\cite{b17a}
\begin{equation}
\mathbf{x}_{IQ} = \alpha_{T} \mathbf{x} + \beta_{T} \mathbf{x^*},
\label{eq12}
\end{equation}
where $\alpha_{T} = \frac{1}{2}\left[ \cos(\frac{\Delta \phi_{T}}{2}) + j\frac{\Delta G_{T}}{2} \sin(\frac{\Delta \phi_{T}}{2})\right]$,
$\beta_{T} = \frac{1}{2}\left[-\frac{\Delta G_{T}}{2} \cos(\frac{\Delta \phi_{T}}{2}) - j\sin(\frac{\Delta \phi_{T}}{2})\right]$, $\mathbf{x}$ is the ideal OTFS transmit vector (without Tx IQI) and $\mathbf{x^*}$, its conjugate, is the image signal. The received vector, assuming an ideal RF chain at the receiver (without Rx IQI) is given by
\begin{equation}
\mathbf{y=Hx_\mathit{IQ}+v}.
\label{eq13}
\end{equation}

{\em Receiver IQI:} Now, consider that the receiver RF chain is also impaired by IQI. Let $\Delta G_{R}$ and $\Delta \phi_{R}$ denote the gain and phase imbalances, respectively, at the receiver. The received vector in the presence of receiver IQI can then be written as 
\begin{equation}
\mathbf{y}_{IQ} = \alpha_{R} \mathbf{y} + \beta_{R} \mathbf{y^*},
\label{eq14}
\end{equation}
where $\alpha_{R} = \frac{1}{2}\left[ \cos(\frac{\Delta \phi_{R}}{2}) + j\frac{\Delta G_{R}}{2} \sin(\frac{\Delta \phi_{R}}{2})\right]$,
$\beta_{R} = \frac{1}{2}\left[-\frac{\Delta G_{R}}{2} \cos(\frac{\Delta \phi_{R}}{2}) - j\sin(\frac{\Delta \phi_{R}}{2})\right]$, $\mathbf{y}$ is the received signal in the absence of receiver IQI and $\mathbf{y^*}$ is the image signal. It is noted that $\alpha_{R}=1$ and $\beta_{R}=0$ in the absence of Rx IQI. Likewise, $\alpha_{T}=1$ and $\beta_{T}= 0$ in the absence of Tx IQI. In the presence of IQI, the image signal causes interference to the desired signal resulting in degraded performance.

\section{Proposed DNN-based OTFS Transceiver}
\label{sec3}
In this section, we present the proposed DNN-based OTFS transceiver which performs the following tasks: $i$) IQI compensation at the transmitter, $ii$) IQI compensation at the receiver, and $iii$) channel training and detection at the receiver. The proposed transceiver architecture is shown in Fig. \ref{fig2}. 

\subsection{Tx IQI compensation}
\label{sec3a}
The Tx IQI compensation is carried out by DNN-1 in Fig. \ref{fig2}, which is a fully-connected DNN with $2|\mathbb{A}|$ input neurons and $2|\mathbb{A}|$ output neurons, where $|\mathbb{A}|$ is the size of the alphabet $\mathbb{A}$. The alphabet $\mathbb{A}$ is given as input to the Tx IQI model in (\ref{eq12}) which generates the alphabet affected by Tx IQI (denoted by $\mathbb{A}_{\mbox{\scriptsize iq}}$). The DNN takes the real and imaginary parts of $\mathbb{A}_{\mbox{\scriptsize iq}}$ and training data as input and generates a compensating alphabet (denoted by $\mathbb{A}_{\mbox{\scriptsize comp}}$), which is fed to the transmit chain. The transmit chain maps $\mathbb{A}$ to $\mathbb{A}_{\mbox{\scriptsize comp}}$ and sends the symbols from $\mathbb{A}_{\mbox{\scriptsize comp}}$ through the chain. This ensures that the alphabet symbols at the transmitter output are pre-compensated and the effect of IQI is nullified. The data used for training the DNN is multiple realizations of ($\mathbb{A}'_{\mbox{\scriptsize iq}}$,$\mathbb{A}'_{\mbox{\scriptsize comp}})$ pairs, where $\mathbb{A}'_{\mbox{\scriptsize iq}}$s are generated using the Tx IQI model for different ($\Delta G_{T}$,$\Delta \phi_{T}$) values, and $\mathbb{A}'_{\mbox{\scriptsize comp}}$s are obtained using the compensation model, given by
\begin{equation}
\begin{bmatrix}
\mathbb{A}'_{\mbox{\scriptsize comp}} \\
\mathbb{A}'^{*}_{\mbox{\scriptsize comp}}
\end{bmatrix} = 
\begin{bmatrix}
\alpha_{T} & \beta_{T} \\
\beta^{*}_{T} & \alpha^{*}_{T}
\end{bmatrix}^{-1}
\begin{bmatrix}
\mathbb{A} \\
\mathbb{A}^{*}
\end{bmatrix}.
\end{equation}

\subsection{Rx IQI compensation}
\label{sec3b}
At the receiver, Rx IQI compensation is done using two DNNs, namely, DNN-2 and DNN-3, as shown in Fig. \ref{fig2}. DNN-2 is a fully connected DNN with $2|\mathbb{A}|$ input neurons and 2 output neurons meant for estimating the gain and phase imbalances. DNN-3 is also a fully connected network with $2MN$ input neurons and $2MN$ output neurons meant for compensating the imbalances. In DNN-2, the Rx IQI impaired alphabet vector (denoted by $\mathbf{s}_{\mbox{\scriptsize iq}}$) is generated by using $\mathbb{A}$ (e.g., BPSK) in the IQI model in (\ref{eq14}). It takes the real and imaginary parts of $\mathbf{s}_{\mbox{\scriptsize iq}}$ and training data as input and gives the estimated values of gain and phase imbalances (denoted by $\Delta \hat{ G}_{R}$ and $\Delta \hat{\phi}_{R}$, respectively) as the output.  The data used for training DNN-2 is multiple realizations of ($\mathbf{s}'_{\mbox{\scriptsize iq}},\mathbf{e}'_{\mbox{\scriptsize est}}$) pairs, where $\mathbf{e}'_{\mbox{\scriptsize est}} = [\Delta G_{R};\Delta \phi_{R}]$ and $\mathbf{s}'_{\mbox{\scriptsize iq}}$s are obtained using different $(\Delta G_{R},\Delta \phi_{R})$ values in (\ref{eq14}). The estimated $\Delta \hat{ G}_{R}$ and $\Delta \hat{\phi}_{R}$ obtained from DNN-2 are subsequently used for training DNN-3 meant for Rx IQI compensation. In DNN-3, the received signal vector $\mathbf{y}$ is given as input to the Rx IQI model in (\ref{eq14}) which generates the received vector affected by IQI (denoted by $\mathbf{y}_{\mbox{\scriptsize iq}}$). It takes the real and imaginary parts of $\mathbf{y}_{\mbox{\scriptsize iq}}$ and training data as input and generates a compensated received vector $\mathbf{y}_{\mbox{\scriptsize comp}}$, which is expected to be same as the vector $\mathbf{y}$ received in the absence of Rx IQI. The Rx IQI compensated vector $\mathbf{y}_{\mbox{\scriptsize comp}}$ is subsequently used for signal detection in DNN-4.
 The training data for DNN-3 consists of multiple realizations of ($\mathbf{y}'_{\mbox{\scriptsize iq}},\mathbf{g}'_{\mbox{\scriptsize est}}$) pairs where $\mathbf{g}'_{\mbox{\scriptsize est}}=[\Delta G_{R};\Delta \phi_{R}]$ and $\mathbf{y}'_{\mbox{\scriptsize iq}}$s are
 generated according to (\ref{eq14}).

\subsection{Channel training and detection}
\label{sec3c}
The task of channel training and detection is carried out by DNN-4 in Fig. \ref{fig2}, which is a fully-connected DNN with $4MN$ input neurons and $MN$ output neurons. The input to the DNN consists of two OTFS frames. The first frame is a pilot frame consisting of pilot symbol(s) and the second frame is a data frame consisting of information symbols. The DD channel is considered to be constant over these two frames. The pilot frame is assumed to consist of one pilot symbol at a fixed pilot location in the DD grid. At the receiver, these two frames are vectorized to form the input to the DNN as $\mathbf{y} = [\mathbf{y}_{\mbox{\scriptsize p}};\mathbf{y}_{\mbox{\scriptsize d}}]$, where $\mathbf{y}_{\mbox{\scriptsize p}}$ and $\mathbf{y}_{\mbox{\scriptsize d}}$ are the received vectors corresponding to the pilot frame and the data frame, respectively. The same pilot frame is used for all the data frames in a spatial coherence interval. The DNN takes the real and imaginary parts of the vector $\mathbf{y}$ as input and recovers the information bits in the data frame. The training data for the DNN consists of multiple realizations of ($\mathbf{y}',\mathbf{x}'$) pairs, where $\mathbf{x}'$s are pseudo-randomly generated transmit vectors, and $\mathbf{y}'$s are obtained using (\ref{eq11}) for different realizations of DD channel coefficients.
 
\section{Simulation results}
\label{sec4}
In this section, we first present the simulated bit error rate (BER) performance of OTFS in the presence of IQI and compare it with that of OFDM. We then illustrate the efficiency of the proposed IQI compensating DNN architectures in mitigating the effects of IQI at the transmitter and receiver. Next, we present the performance of channel training and detection DNN and compare it with those of ML detection and MMSE detection using impulse based channel estimation. Finally, we present the combined performance of all the four DNNs in the presence of both Tx and Rx IQI.

We consider the OTFS system parameters listed in Table \ref{tab} for all the simulations.

\begin{table}
\vspace{3mm}
\centering
\footnotesize
\begin{tabular}{|l|c|} 
\hline
\textbf{Parameter} & \textbf{Value}\\ [0.3ex] 
\hline\hline
Frame size $(M,N)$ & $(4,4)$\\ 
\hline
Carrier frequency (GHz) & $4$\\
\hline
Subcarrier spacing (kHz) & $3.75$\\
\hline
No. of DD channel paths $(P)$ & $4$\\
\hline
Delay-Doppler profile $(\tau_i, \nu_i)$ & $(0,\frac{1}{NT}),(\frac{1}{M \Delta f},\frac{1}{NT})$,\\
 & $(\frac{2}{M \Delta f},\frac{1}{NT}),(\frac{3}{M \Delta f},\frac{1}{NT})$\\
\hline
Delay power profile & Uniform\\
\hline
\end{tabular}
\vspace{2mm}
\caption{OTFS system parameters.}
\vspace{-6mm}
\label{tab}
\end{table}

\subsection{BER performance of OTFS and OFDM with IQI}
\label{sec4a}
\begin{figure*}[t]
\centering
\subfloat[\centering $\Delta \phi_{T}$ sensitivity]{{\includegraphics[width=8cm,height=4.75cm]{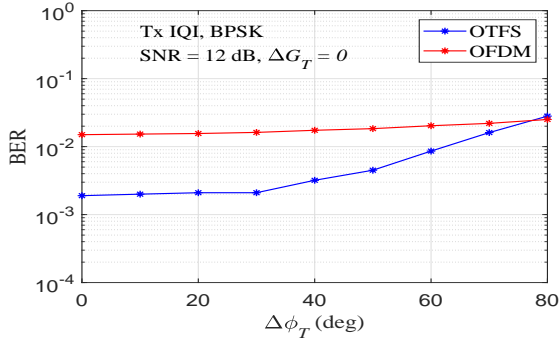}}}
\subfloat[\centering $\Delta G_{T}$ sensitivity]{{\includegraphics[width=8cm,height=4.75cm]{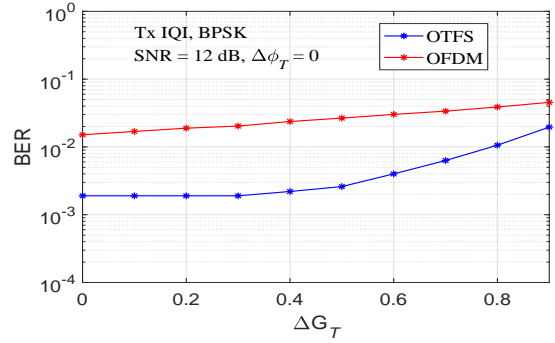}}}
\qquad
\subfloat[\centering $\Delta \phi_{R}$ sensitivity]{{\includegraphics[width=8cm,height=4.75cm]{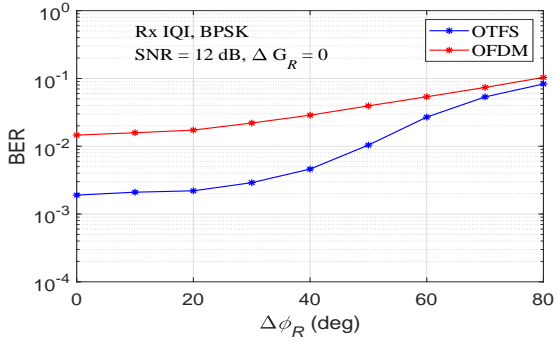}}}
\subfloat[\centering $\Delta G_{R}$ sensitivity]{{\includegraphics[width=8cm,height=4.75cm]{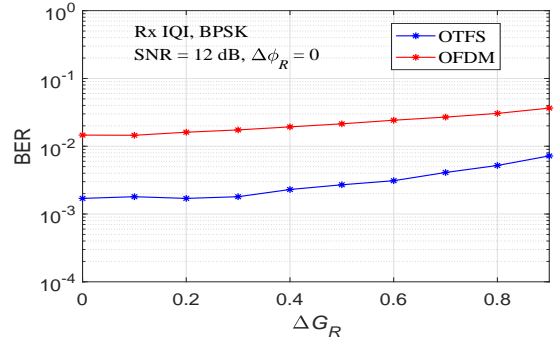}}}
\caption{Sensitivity of OTFS and OFDM to Tx and Rx IQI.} 
\label{fig3}
\vspace{-4mm}
\end{figure*}

Here, the effect of Tx IQI (assuming ideal receiver) and Rx IQI (assuming ideal transmitter) on OTFS is analyzed. Figure \ref{fig3} shows the BER performance comparison between OTFS and OFDM as a function of IQI parameters $\Delta G$ and $\Delta \phi$ at the transmitter and receiver. The gain imbalance $\Delta G$ is varied by considering $\Delta \phi = 0$. Similarly, $\Delta \phi$ is varied by fixing the value of $\Delta G$ at $0$. The signal-to-noise ratio (SNR) considered is 12 dB. It can be seen that BER performance of OTFS is better compared to OFDM in the presence of Tx IQI and Rx IQI. In case of OTFS, the BER is in the order of $10^{-3}$ while it is in the order of $10^{-2}$ in OFDM. OTFS is more resilient to IQI when $\Delta G \in [0,0.3]$ and $\Delta \phi \in [0^{\circ},20^{\circ}]$. Though OTFS becomes less sensitive to IQ imbalance for high values of gain and phase imbalances, its BER is still better than that of OFDM.

\vspace{-1.5mm}
\subsection{Tx IQI compensation}
\label{sec4b}
\begin{figure}[H]
\centering
\includegraphics[width=8.5cm, height=6cm]{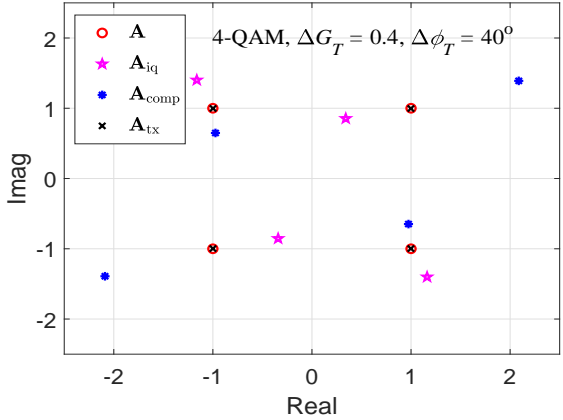}
\caption{Illustration of ideal 4-QAM alphabet ($\mathbb{A}$), Tx IQI impaired alphabet ($\mathbb{A}_{\mbox{\scriptsize iq}})$, compensating alphabet ($\mathbb{A}_{\mbox{\scriptsize comp}}$), and transmit alphabet ($\mathbb{A}_{\mbox{\scriptsize tx}}$) in DNN-1 for $\Delta G=0.4$, $\Delta \phi = 40^\circ$.}
\label{fig4}
\vspace{-4mm}
\end{figure}

\begin{figure}[H]
\vspace{-2mm}
\centering
\includegraphics[width=8.5cm, height=6cm]{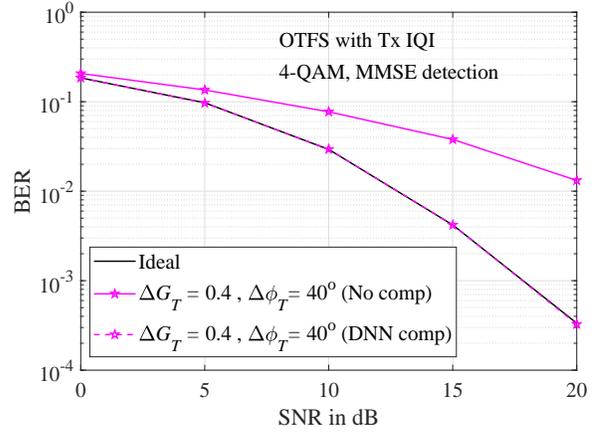}
\caption{BER performance of OTFS with Tx IQI compensation using DNN-1 for 4-QAM.}
\label{fig5}
\vspace{-4mm}
\end{figure}

Here, we consider the Tx IQI compensation performance of DNN-1 assuming that the receiver RF chain is ideal. Table \ref{tab1} shows the parameters used in DNN-1 for BPSK, 4-QAM, and 16-QAM. The DNN-1 for 4-QAM has five layers with one input layer, three hidden layers, and one output layer. The activation function used for hidden layers is Tanh activation. Linear activation function is used for the output layer so that the real-valued output is not confined between any range. Predicting a real-valued output is a regression problem and mean-squared error loss function is chosen because it is sensitive to large deviations and minimizes the average squared difference between the DNN output and the expected output. The DNN-1 architecture for 4-QAM is as follows. \newline
{\em {\bfseries DNN-1 for 4-QAM:}} 
{\em Input $\rightarrow$ 8 $\rightarrow$ Tanh $\rightarrow$ 64 $\rightarrow$ Tanh $\rightarrow$ 32 $\rightarrow$ Tanh $\rightarrow$ 16 $\rightarrow$ Tanh $\rightarrow$ 8 $\rightarrow$ Linear.}
\newline
In Fig. \ref{fig4}, we illustrate the effectiveness of the Tx IQI compensation achieved by DNN-1. It shows the ideal 4-QAM alphabet $\mathbb{A}$, the Tx IQI impaired alphabet $\mathbb{A}_{\mbox{\scriptsize iq}}$, the compensating alphabet $\mathbb{A}_{\mbox{\scriptsize comp}}$, and the transmit alphabet $\mathbb{A}_{\mbox{\scriptsize tx}}$ for $\Delta G_{T} = 0.4$ and $\Delta \phi_{T} = 40^\circ$. From Fig. \ref{fig4}, it can be seen that the transmitted symbols after compensation are almost same as the corresponding symbols from the ideal alphabet $\mathbb{A}$. This shows that the DNN-1 is effective in compensating the Tx IQI.  This effectiveness translates into very good BER performance as shown in Fig. \ref{fig5}. Figure \ref{fig5} shows three BER plots as a function of SNR for the following three cases: 1) ideal Tx with no IQI, 2) Tx with IQI and with no IQI compensation, and 3) Tx with IQI and with proposed IQI compensation. The symbols are detected using MMSE detection. It can be seen that, though Tx IQI significantly degrades the BER performance, the performance achieved by the proposed Tx IQI compensation using DNN-1 is almost the same as the ideal BER performance achieved in the absence of IQI.

\begin{table}
\centering
\footnotesize
\vspace{3mm}
\begin{tabular}{|l|c|c|c|} 
\hline
Parameters & BPSK-DNN & 4QAM-DNN & 16QAM-DNN\\ [0.3ex] 
\hline\hline
No. of input neurons & $2|\mathbb{A}|=4$ & $2|\mathbb{A}|=8$ & $2|\mathbb{A}|=32$ \\ 
\hline
No. of output neurons & $2|\mathbb{A}|=4$ & $2|\mathbb{A}|=8$ & $2|\mathbb{A}|=32$ \\
\hline
No. of hidden layers & 4 & 3 & 3 \\
\hline
Hidden layer activation & Tanh & Tanh & Tanh \\
\hline
Output layer activation & Linear & Linear & Linear \\
\hline
Optimization & Adam & Adam & Adam\\
\hline
Loss function & MSE & MSE & MSE\\
\hline
No. of training examples & 1000 & 1000 & 1000\\
\hline
No. of epochs & 5000 & 5000 & 5000 \\
\hline
Batch size & 5 & 5 & 5\\
\hline
\end{tabular}
\vspace{2mm}
\caption{Parameters of Tx IQI compensation DNN-1.}
\vspace{-5mm}
\label{tab1}
\end{table} 

\begin{figure*}
\centering
\subfloat[\centering BPSK]{{\includegraphics[width=8cm,height=5cm]{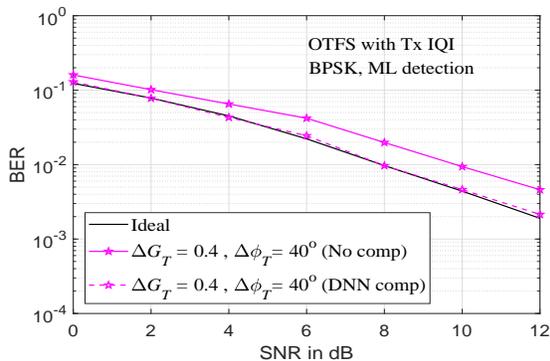} }}
\subfloat[\centering 16-QAM]{{\includegraphics[width=8cm,height=5cm]{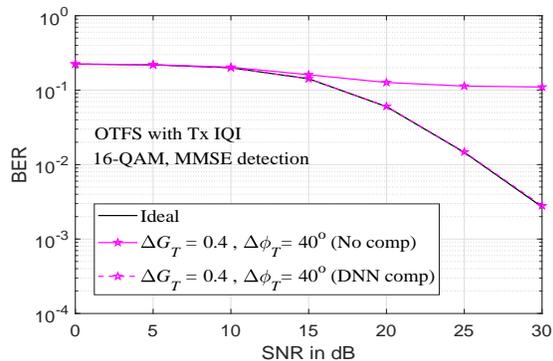} }}
\caption{BER performance of OTFS with Tx IQI compensation using DNN-1 for BPSK and 16-QAM.}
\label{fig6}
\vspace{-2mm}
\end{figure*}

A similar restoration of BER performance is achieved by DNN-1 in the case of BPSK and 16-QAM as well, as can be seen in Fig. \ref{fig6}. ML and MMSE detection are used for BPSK and 16-QAM, respectively. The DNN architectures used for BPSK and 16-QAM are as follows.\newline
{\em {\bfseries DNN-1 for BPSK:}} 
{\em Input $\rightarrow$ 4 $\rightarrow$ Tanh $\rightarrow$ 64 $\rightarrow$ Tanh $\rightarrow$ 32 $\rightarrow$ Tanh $\rightarrow$ 16 $\rightarrow$ Tanh $\rightarrow$ 8 $\rightarrow$ Tanh $\rightarrow$ 4 $\rightarrow$ Linear.}\newline
{\em {\bfseries DNN-1 for 16-QAM:}}
{\em Input $\rightarrow$ 32 $\rightarrow$ Tanh $\rightarrow$ 256 $\rightarrow$ Tanh $\rightarrow$ 128 $\rightarrow$ Tanh $\rightarrow$ 64 $\rightarrow$ Tanh $\rightarrow$ 32 $\rightarrow$ Linear.}\newline
From both Figs. \ref{fig5} and \ref{fig6}, we observe that the proposed DNN-1 has effectively nullified the effect of Tx IQI and achieved near-ideal performance.

\subsection{Rx IQI compensation}
\label{sec4c}
Here, we consider the performance of DNN-2 and DNN-3 for the estimation and compensation of Rx IQI in the absence of Tx IQI. First, the Rx IQI parameters, $\Delta \hat{G}_{R}$ and $\Delta \hat{\phi}_{R}$, are estimated using DNN-2 using a test alphabet $\mathbb{A}$ as the input. The DNN-3 is trained separately for BPSK, 4-QAM, and 16-QAM alphabets. The IQ impaired received vector at the receiver is passed through DNN-3 and the output obtained is used for detection using conventional methods like ML detection and MMSE detection. Table \ref{tab2} shows the parameters used in DNN-2 and DNN-3.
 The DNN-2 and DNN-3 architectures used are as follows. \newline
{\em {\bfseries DNN-2:}} 
{\em Input $\rightarrow$ 4 $\rightarrow$ Tanh $\rightarrow$ 8 $\rightarrow$ Tanh  $\rightarrow$ 2 $\rightarrow$ Linear.} \newline
{\em {\bfseries DNN-3:}} 
{\em Input $\rightarrow$ 32 $\rightarrow$ Tanh $\rightarrow$ 64 $\rightarrow$ Tanh  $\rightarrow$ 32 $\rightarrow$ Linear.}\newline
\begin{table}[H]
\centering
\footnotesize
\begin{tabular}{|l|c|c|c|} 
\hline
Parameters & DNN-2 & DNN-3\\ [0.3ex] 
\hline\hline
No. of input neurons & $2|\mathbb{A}|=4$ & $2MN=32$\\ 
\hline
No. of output neurons & $2$ & $2MN=32$ \\
\hline
No. of hidden layers & 1 & 1\\
\hline
Hidden layer activation & Tanh & Tanh\\
\hline
Output layer activation & Linear & Linear\\
\hline
Optimization & Adam & Adam\\
\hline
Loss function & MSE & MSE\\
\hline
No. of training examples & 1000 & 50000\\
\hline
No. of epochs & 500 & 500\\
\hline
Batch size & 5 & 50\\
\hline
\end{tabular}
\vspace{2mm}
\caption{Parameters of Rx IQI DNN-2 and DNN-3.}
\vspace{-2mm}
\label{tab2}
\end{table}

\begin{figure*}
\centering
\subfloat[\centering BPSK]{{\includegraphics[width=6cm,height=5cm]{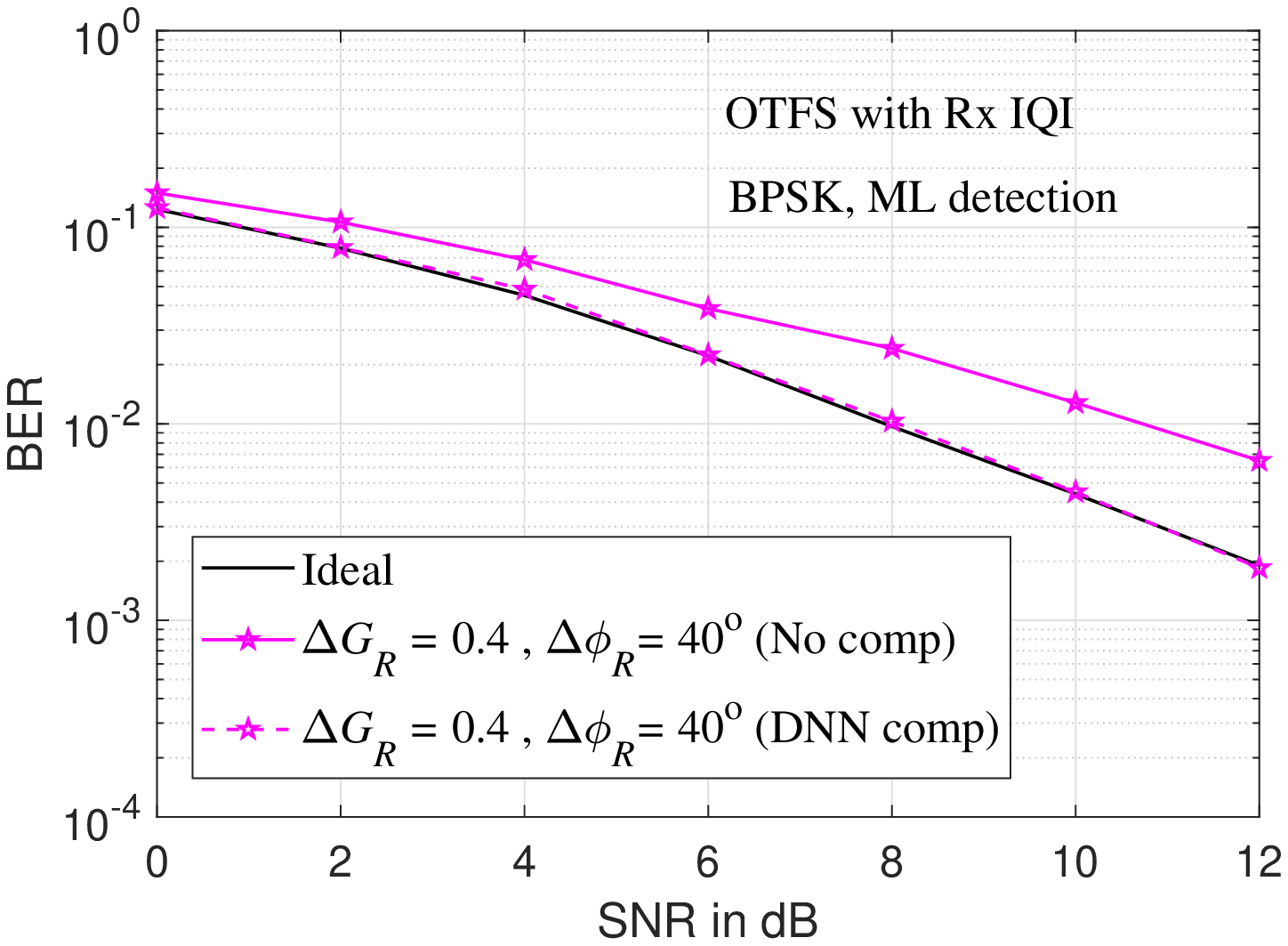}}}
\subfloat[\centering 4-QAM]{{\includegraphics[width=6cm,height=5cm]{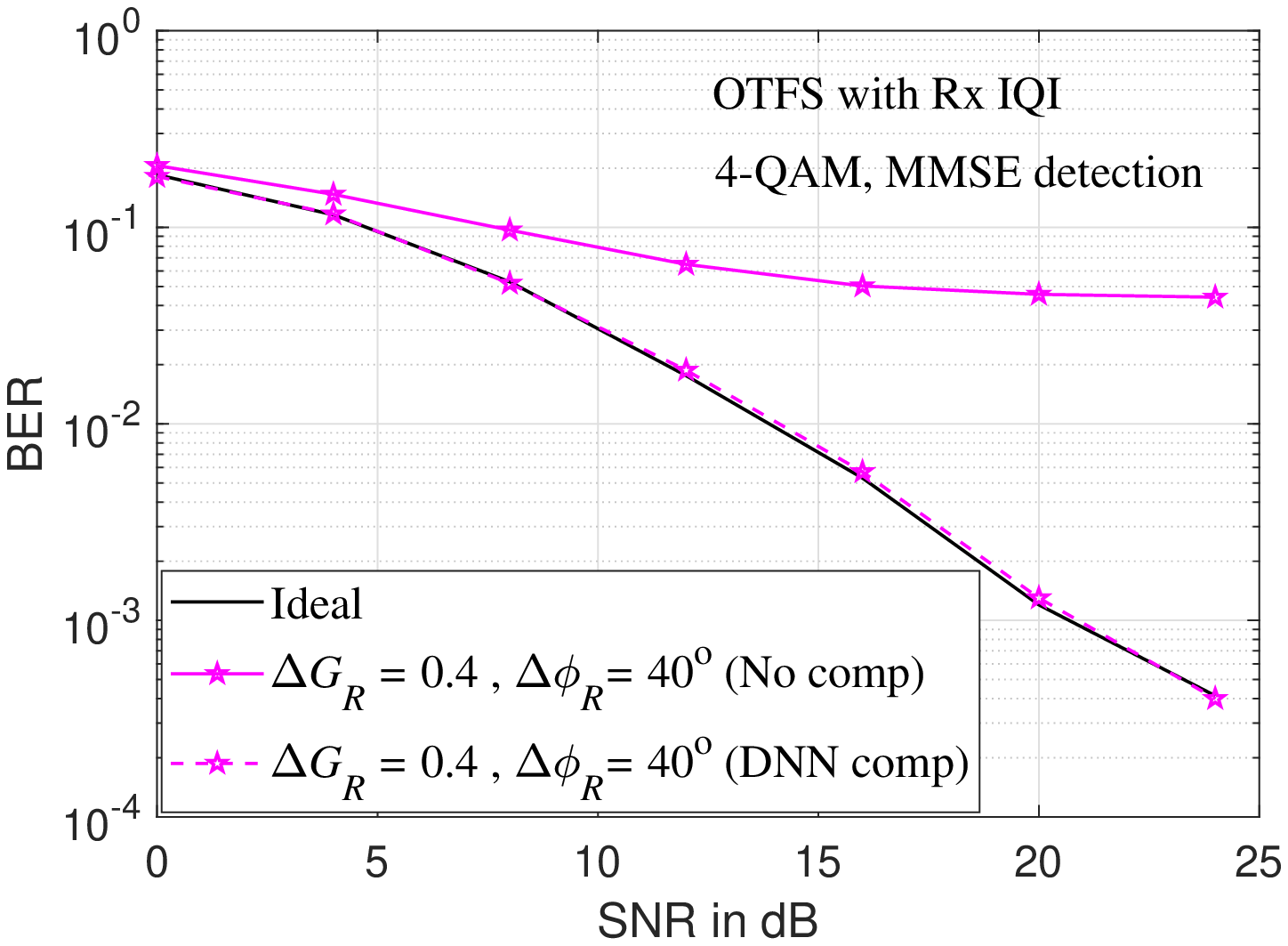}}}
\subfloat[\centering 16-QAM]{{\includegraphics[width=6cm,height=5cm]{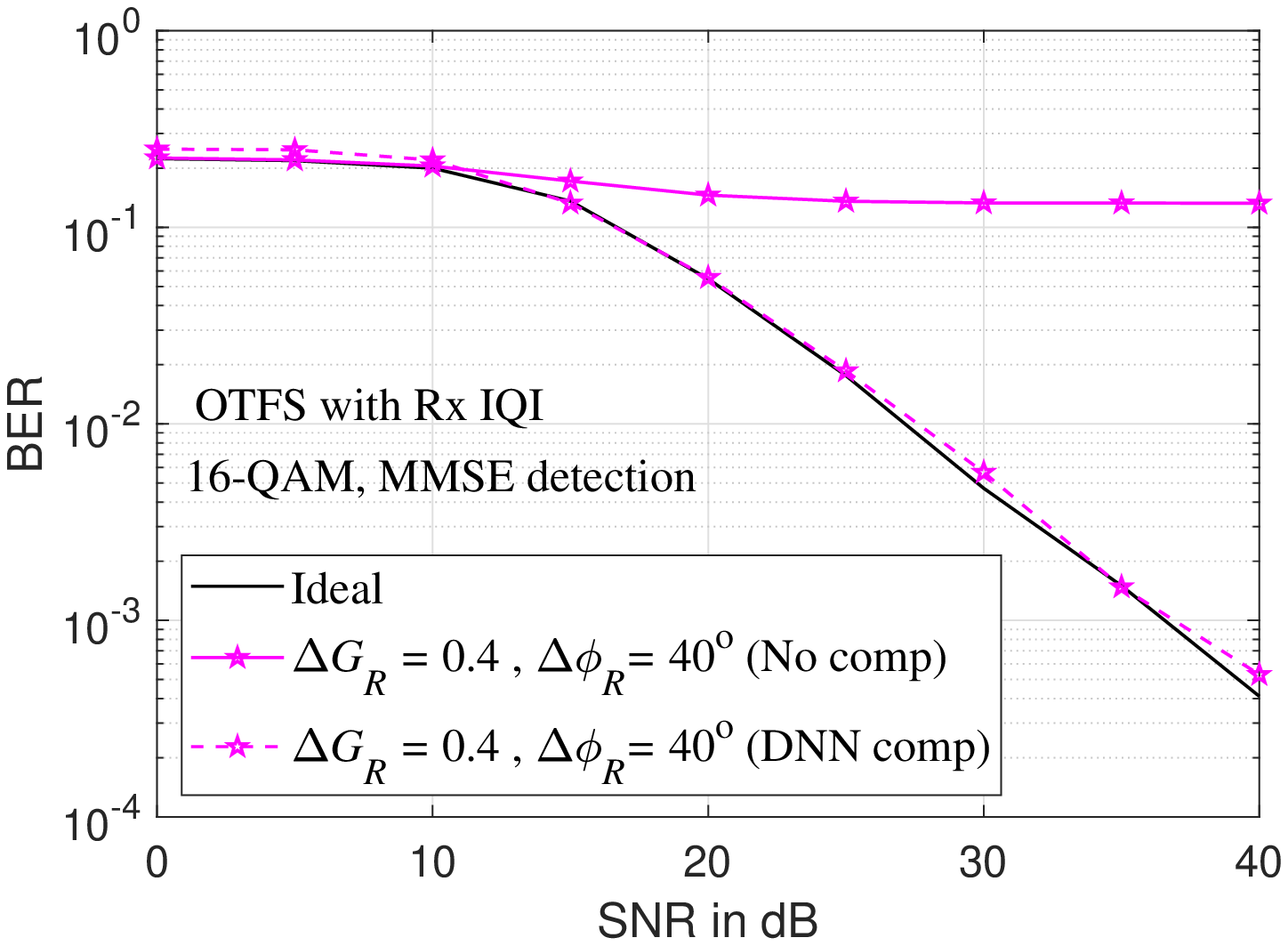}}}
\caption{BER performance of OTFS with Rx IQI estimation and compensation DNN-2 and DNN-3 for BPSK, 4-QAM, and 16-QAM.}
\label{fig7}
\vspace{-4mm}
\end{figure*}

Figures \ref{fig7}(a), (b), and (c) show the combined BER performance of DNN-2 and DNN-3 for BPSK, 4-QAM and 16-QAM, respectively. The values of gain and phase imbalances considered are $\Delta G_{R} = 0.4$ and $\Delta \phi_{R} = 40^{\circ}$. In all the three cases, it can be seen that the performance achieved by the proposed DNN-2 and DNN-3 in the presence of Rx IQI is almost same as the BER performance of an ideal transceiver without IQI. This shows that the proposed Rx IQI DNNs are able to nullify the degrading effects caused by Rx IQI.

\subsection{Channel training and detection}
\label{sec4d}
Here, we demonstrate the performance of DNN-4 for channel training and detection in OTFS and compare it with the performance of conventional methods in terms of BER. An OTFS system with $M=N=4$ and BPSK is considered. The DNN-4 has four layers. The hidden layers use ReLU activation and the output layer uses Sigmoid activation. The output of the DNN has values in the interval [0,1] due to the Sigmoid activation at the output layer. The transmitted bits are recovered by thresholding the output at 0.5. Table \ref{tab3} shows the parameters used in DNN-4. The architecture used in DNN-4 is as follows.\newline
{\em {\bfseries DNN-4:}}
{\em Input $\rightarrow$ 64 $\rightarrow$ ReLU $\rightarrow$ 256 $\rightarrow$ ReLU  $\rightarrow$ 64 $\rightarrow$ ReLU $\rightarrow$ 16 $\rightarrow$ Sigmoid.}

\begin{table}
\centering
\footnotesize
\vspace{3mm}
\begin{tabular}{|l|c|} 
\hline
Parameters & DNN-4\\ [0.3ex] 
\hline\hline
No. of input neurons & $4MN=64$\\ 
\hline
No. of output neurons & $MN=16$ \\
\hline
No. of hidden layers & 2 \\
\hline
Hidden layer activation & ReLU\\
\hline
Output layer activation & Sigmoid\\
\hline
Training data SNR & $10$ dB \\
\hline
Training pilot SNR & $10$ dB\\
\hline
Optimization & Adam\\
\hline
Loss function & MSE\\
\hline
No. of training examples & 200000 \\
\hline
No. of epochs & 500 \\
\hline
Batch size & 500 \\
\hline
\end{tabular}
\vspace{2mm}
\caption{Parameters of channel training and detection DNN-4.}
\vspace{-2mm}
\label{tab3}
\end{table}
  
\begin{figure}[t]
\centering
\includegraphics[width=8.5cm, height=6cm]{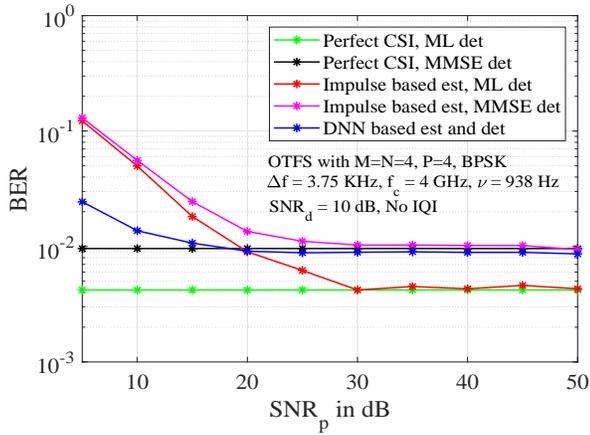}
\caption{BER performance of DNN-4 as a function of pilot SNR.}
\label{fig8}
\vspace{-2mm}
\end{figure}

Figure \ref{fig8} shows the BER performance of DNN-4 as a function of pilot SNR ($SNR_p$). The BER is computed at a data SNR ($SNR_d$) of 10 dB. The DNN-4 is trained at $SNR_p = 10$ dB and $SNR_d = 10$ dB. The DNN-4 once trained can be used over a spatial coherence interval. The performance of DNN-4 is compared with the BER performance of ML detection and MMSE detection which use the channel estimated using impulse based channel estimation \cite{mimo_otfs}. The BER with perfect channel knowledge is also shown for comparison. It can be seen that DNN-4 performs better than the other methods at pilot SNR lower than 20 dB. Beyond that, it performs worse than ML detection but performs slightly better than MMSE detection. Also, it can be noticed that the DNN trained at a particular pilot SNR performs well when tested at a different pilot SNR.

\subsection{Combined performance all DNNs}
\label{sec4e}
\begin{figure*}
\centering
\subfloat[\centering $SNR_p = 10$ dB]{{\includegraphics[width=6cm,height=5cm]{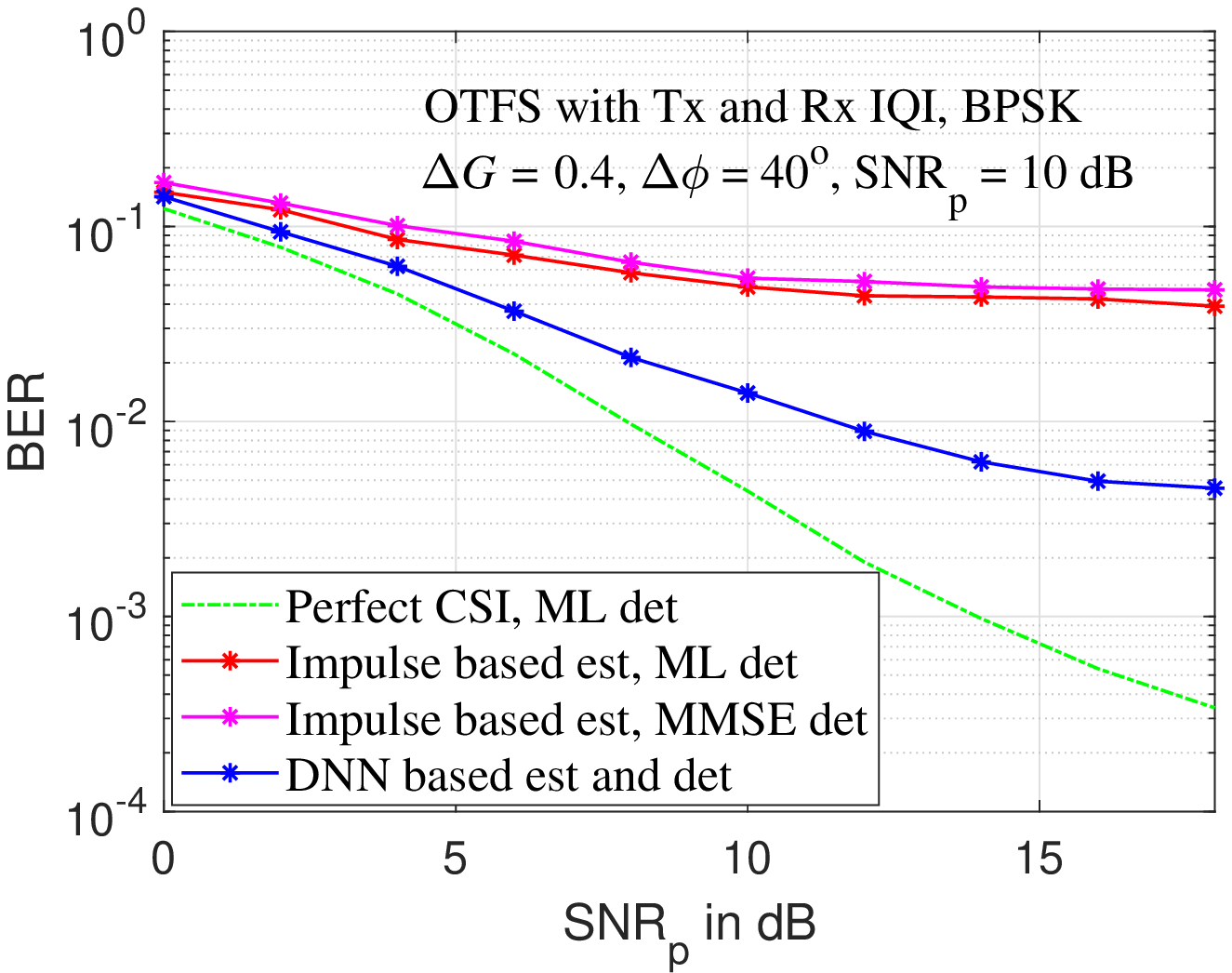}}}
\subfloat[\centering $SNR_p = 20$ dB]{{\includegraphics[width=6cm,height=5cm]{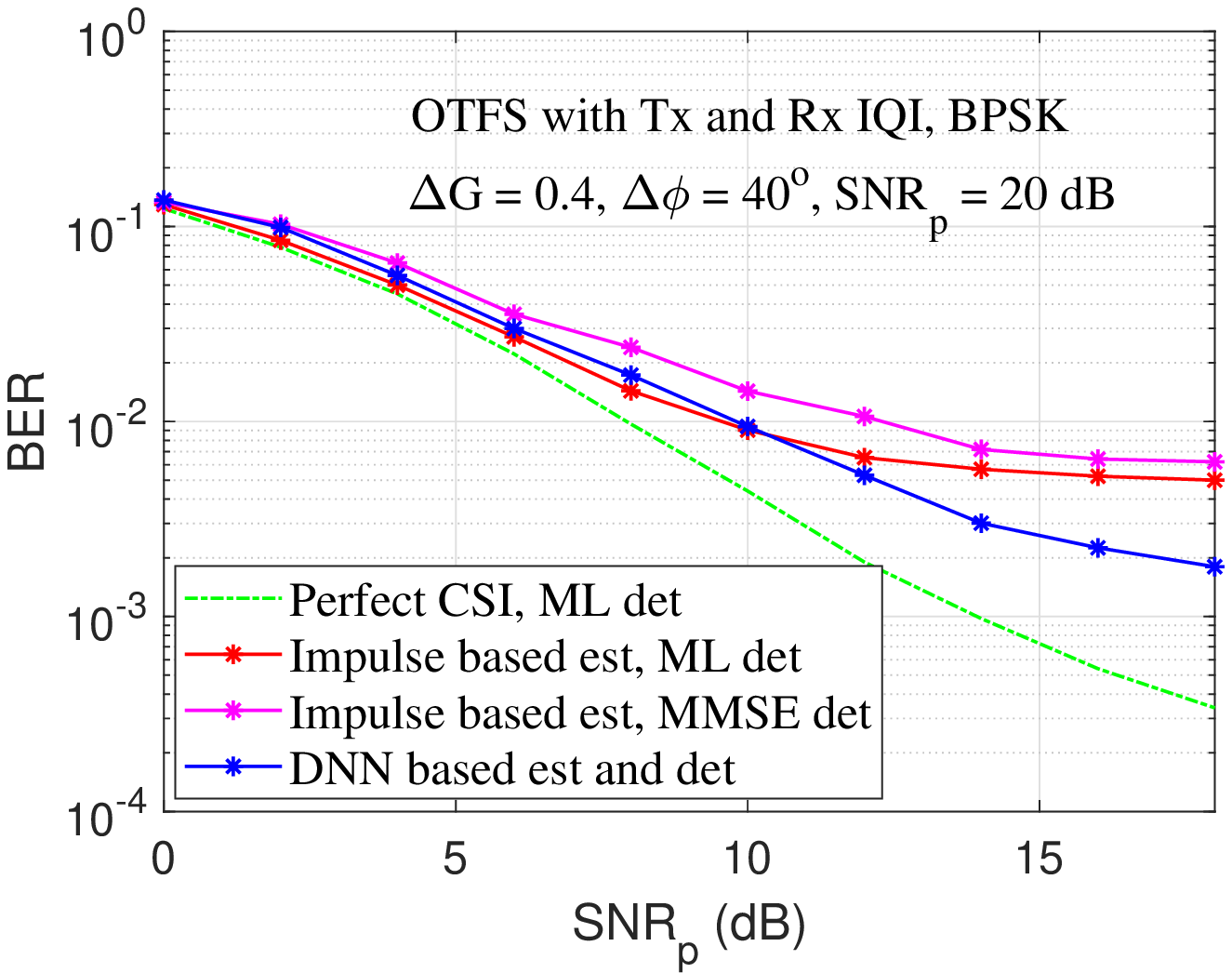}}}
\subfloat[\centering $SNR_p = 30$ dB]{{\includegraphics[width=6cm,height=5cm]{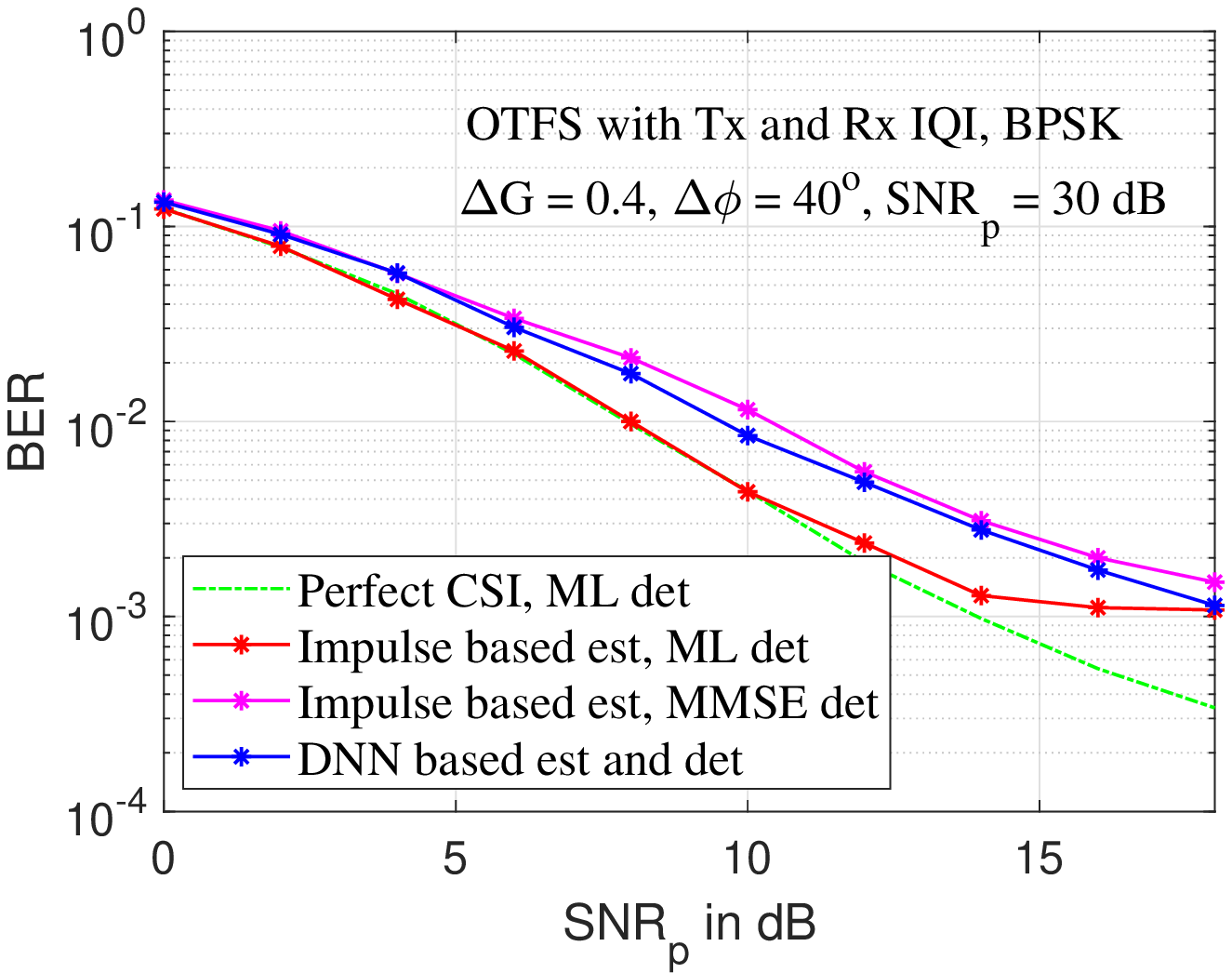}}}
\caption{BER performance of OTFS using the complete DNN-based transceiver with Tx \& Rx IQI compensation and DD channel training/detection at different pilot SNRs.}
\label{fig9}
\end{figure*}

Finally, in Fig. \ref{fig9} we show the BER performance of OTFS using the complete DNN-based transceiver architecture in Fig. \ref{fig2} with Tx \& Rx IQI compensation and DD channel training/detection. After IQI compensation using DNN-1, DNN-2, and DNN-3, BER is compared between $i$) impulse based channel estimation and ML detection, $ii$) impulse based channel estimation and MMSE detection, and $iii$) DNN-based channel training and detection using DNN-4. The performance of ML detection with perfect channel knowledge is also shown. The comparison is done at $SNR_p = 10$ dB, 20 dB, 30 dB. It can be seen that, when the pilot SNR is small, both ML detection and MMSE detection methods show degradation in the BER performance due to increased error in the impulse based estimation of the channel coefficients. The DNN based
detection outperforms the other detection methods at $SNR_p = 10$ dB and $SNR_p = 20$ dB because of effective channel training. At $SNR_p = 30$ dB, ML detection with impulse based estimation gives the best performance. Also, the performance of DNN-4 is comparable to that of MMSE detection with impulse based channel estimation. 

\section{Conclusion}
\label{sec5}
We proposed an integrated DNN-based OTFS architecture to carry out DD channel training, detection, and IQI compensation tasks in OTFS transceivers. The proposed transceiver used a single DNN at the receiver to learn the DD channel over a spatial coherence interval and also detect the information symbols in an OTFS frame. The proposed transceiver also learnt the IQ imbalances effectively and compensated them. The Tx IQI compensation DNN at the transmitter learnt and provided a compensating modulation alphabet to pre-rotate the modulation symbols before sending through the transmitter. The Rx IQI imbalance compensation was realized using two DNNs at the receiver, one DNN for explicit estimation of receive gain and phase imbalances and another DNN for compensation. Simulation results showed very good performance for the proposed DNN-based OTFS transceiver suggesting that the DNN-based approach is  attractive for the design of practical transceivers. 

\begin{thebibliography}{00}
\bibitem{b1} R. Hadani et al., ``Orthogonal time frequency space modulation,'' in {\em Proc. IEEE WCNC'2017}, Mar. 2017, pp. 1-6.

\bibitem{b2} 
R. Hadani et al., ``Orthogonal time frequency space modulation,'' available online: arXiv:1808.00519v1 [cs.IT] 1 Aug 2008.

\bibitem{b3} 
G. D. Surabhi, R. M. Augustine, and A. Chockalingam, ``On the diversity of uncoded OTFS modulation in doubly-dispersive channels,'' {\em IEEE Trans. Wireless Commun.}, vol. 18, no. 6, pp. 3049-3063, Jun. 2019.

\bibitem{b4}
F. Wiffen, L. Sayer, M. Z. Bocus, A. Doufexi, and A. Nix, ``Comparison of OTFS and OFDM in ray launched sub-6 GHz and mmWave line-of-sight mobility channels,'' in {\em Proc. IEEE PIMRC'2018}, Sep. 2018, pp. 73-79.

\bibitem{b5}
S. K. Mohammed, ``Derivation of OTFS modulation from first principles,'' {\em IEEE Trans. Veh. Tech.}, early access in IEEE Xplore, DOI: 10.1109/TVT.2021.3069913, 31 Mar. 2021.

\bibitem{b6} 
P. Raviteja, K. T. Phan, Y. Hong, and E. Viterbo, ``Interference cancellation and iterative detection for orthogonal time frequency space modulation,'' {\em IEEE Trans. Wireless Commun.}, vol. 17, no. 10, pp. 6501-6515, Oct. 2018.

\bibitem{b7} 
K. R. Murali and A. Chockalingam, ``On OTFS modulation for high-Doppler fading channels,'' in {\em Proc. ITA'2018}, Feb. 2018, pp. 1-10.

\bibitem{b8} 
R. Hadani and A. Monk, ``OTFS: a new generation of modulation addressing the challenges of 5G,'' available online: arXiv:1802.02623v1 [cs.IT] 7 Feb 2018. 

\bibitem{mimo_otfs}
M. K. Ramachandran and A. Chockalingam, ``MIMO-OTFS in high-Doppler fading channels: signal detection and channel estimation,'' in {\em Proc. IEEE GLOBECOM'2018}, Dec. 2018.

\bibitem{b9}
P. Raviteja, K. T. Phan, and Y. Hong, ``Embedded pilot-aided channel estimation for OTFS in delay–Doppler channels,'' {\em IEEE Trans. Veh. Tech.}, vol. 68, no. 5, pp. 4906-4917, May 2019.

\bibitem{b10}
H. Ye, G. Y. Li, and B. Juang, ``Power of deep learning for channel estimation and signal detection in OFDM systems,'' {\em IEEE Wireless Commun. Lett.}, vol. 7, no. 1, pp. 114-117, Feb. 2018.

\bibitem{b11} 
S. Dorner, S. Cammerer, J. Hoydis, and S. t. Brink, ``Deep learning based communication over the air,'' {\em IEEE J. Sel. Topics in Signal Process.}, vol. 12, no. 1, pp. 132-143, Feb. 2018. 

\bibitem{b12}
T. O\textquoteright Shea and J. Hoydis, ``An introduction to deep learning for the physical layer,'' {\em IEEE Trans. Cognitive Commun. and Netw.}, vol. 3, pp. 563-575, Dec. 2017.

\bibitem{b13}
Y. Jiang, H. Kim, H. Asnani, S. Kannan, S. Oh, and P. Viswanath, ``LEARN codes: inventing low-latency codes via recurrent neural networks,'' in {\em Proc. IEEE ICC'2019}, Jul. 2019.

\bibitem{b14} 
H. Kim, Y. Jiang, R. Rana, S. Kannan, and P. Viswanath, ``Communication algorithms via deep learning,'' in {\em Proc. ICLR'2018}, Apr.-May 2018, pp. 1-17.

\bibitem{b15} 
N. Farsad and A. Goldsmith, ``Neural network detection of data sequences in communication systems,'' {\em IEEE Trans. Signal Process.}, vol. 66, no. 21, pp. 5663-5678, Sep. 2018. 

\bibitem{b16} 
T. V. Luong, Y. Ko, N. A. Vien, D. H. N. Nguyen, and M. Matthaiou, ``Deep learning-based detector for OFDM-IM,'' {\em IEEE Wireless Commun. Lett.}, vol. 8, no. 4, pp. 1159-1162, Aug. 2019. 

\bibitem{b16a}
T. J. Rouphael, {\em Wireless Receiver Architectures and Design: Antenna, RF, Synthesizers, Mixed Signal and Digital Signal Processing}, Academic Press, 2014.

\bibitem{b17}
T. C. W. Schenk, {\em RF Impairments in High-Rate Wireless Systems}, Springer, 2008.

\bibitem{b17a}
L. Smaini, {\em RF Analog Impairments Modeling for Communication Systems Simulation: Application to OFDM-based Transceivers}, John-Wiley \& Sons, 2012.

\bibitem{b19}
R. Marsalek, J. Blumenstein, D. Schützenhöfer, and M. Pospisil, ``OTFS modulation and influence of wideband RF impairments measured on a 60 GHz testbed,'' in {\em Proc. IEEE  SPAWC'2020}, May 2020, pp. 1-5.

\bibitem{b20}
S. G. Neelam and P. R. Sahu, ``Error performance of OTFS in the presence of IQI and PA nonlinearity,'' {\em NCC'2020}, Feb. 2020, pp. 1-6.

\bibitem{b21}
J. Tubbax, B. Come, L. Van der Perre, S. Donnay, M. Moonen, and H. De Man, ``Compensation of transmitter IQ imbalance for OFDM systems,'' in {\em Proc. IEEE ICASSP'2004}, May 2004, pp. 325-328.

\end{thebibliography}
\end{document}